\def\MPL #1 #2 #3 {Mod.~Phys.~Lett.~{\bf#1},\  #2 (#3)}
\def\NPB #1 #2 #3 {Nucl.~Phys.~{\bf#1},\  #2 (#3)}
\def\PLB #1 #2 #3 {Phys.~Lett.~{\bf#1},\  #2 (#3)}
\def\PR #1 #2 #3 {Phys.~Rep.~{\bf#1},\ #2 (#3)}
\def\PRD #1 #2 #3 {Phys.~Rev.~{\bf#1},\  #2 (#3)}
\def\PRL #1 #2 #3 {Phys.~Rev.~Lett.~{\bf#1},\  #2 (#3)}
\def\RMP #1 #2 #3 {Rev.~Mod.~Phys.~{\bf#1},\  #2 (#3)}
\def\ZP #1 #2 #3 {Z.~Phys.~{\bf#1},\  #2 (#3)}
\def\IJMP #1 #2 #3 {Int.~J.~Mod.~Phys.~{\bf#1},\  #2 (#3)}
\def\softvisible{{ soft/visible}}
\def\ptmin{p_T^{\gam~{\rm min}}}
\def\thetamin{\theta_{\rm min}}
\def\thetae{\theta_e}
\def\thetagam{\theta_\gam}
\def\emax{E^{\rm max}}
\def\mtest{m_{\rm test}}
\def\wtil{\widetilde}
\def\mchitil{m_{\tilde \chi}}
\def\dmchi{\Delta m_{\tilde \chi_1}}
\def\mzstar{m_{Z^\star}}

\def\etal{{\it et al.}}
\def\epem{e^+e^-}

\def\taup{\tau^+}
\def\taum{\tau^-}

\def\lsim{\mathrel{\raise.3ex\hbox{$<$\kern-.75em\lower1ex\hbox{$\sim$}}}}
\def\gsim{\mathrel{\raise.3ex\hbox{$>$\kern-.75em\lower1ex\hbox{$\sim$}}}}
\def\@versim#1#2{\vcenter{\offinterlineskip
        \ialign{$\m@th#1\hfil##\hfil$\crcr#2\crcr\sim\crcr } }}

\def\slash#1{#1\hskip-6pt/\hskip2pt}

\def\etmiss{\slash E_T}

\def\ie{{\it i.e.}}

\def\gam{\gamma}

\def\nsd{N_{SD}}
\def\anti{\overline}
\def\pbi{~{\rm pb}^{-1}}
\def\fbi{~{\rm fb}^{-1}}

\def\mev{\,{\rm MeV}}
\def\gev{\,{\rm GeV}}
\def\tev{\,{\rm TeV}}

\def\wt{\widetilde}

\def\rta{\rightarrow}
\def\mhalf{m_{1/2}}
\def\gl{\wt g}
\def\mgl{m_{\gl}}

\def\tanb{\tan\beta}

\def\mz{m_Z}

\def\mgut{M_U}

\def\cnone{\wt\chi^0_1}
\def\cntwo{\wt\chi^0_2}
\def\snu{\wt\nu}
\def\snubar{\anti{\snu}}
\def\msnu{m_{\snu}}
\def\mcnone{m_{\cnone}}
\def\mcntwo{m_{\cntwo}}

\def\cpone{\wt \chi^+_1}
\def\cmone{\wt \chi^-_1}
\def\cpmone{\wt \chi^{\pm}_1}
\def\mcpone{m_{\cpone}}
\def\mcpmone{m_{\cpmone}}
\def\stau{\wt\tau}
\def\stauone{\wt \tau_1}
\def\mstauone{m_{\stauone}}

\documentstyle[12pt,equations]{article}
\def\ie{{\it i.e.}}
\def\etal{{\it et al.}}
\def\9{\phantom 0}      %%% for lining up numbers in columns
\renewcommand\linebreak{\unskip\break} %% breaks line & still justifies
\begin{document}
\input psfig.sty
\newlength{\captsize} \let\captsize=\small % use \let\normalsize=\captsize
\newlength{\captwidth}                     % just before \caption{ ...
%%%%%%%%%%%%%%%%%%%%%%%%%%%%%%%%%%%%%%%%%%%%%%%%%%%%%%%%%%%%%%

%\preprint{
%
\font\fortssbx=cmssbx10 scaled \magstep2
\hbox to \hsize{
%
%\special{psfile=uwlogo.ps
% hscale=8000 vscale=8000
% hoffset=-12 voffset=-2}
%\hskip.5in \raise.1in
%
$\vcenter{
\hbox{\fortssbx University of California - Davis}
%\hbox{\fortssbx University of Wisconsin - Madison}
}$
\hfill
$\vcenter{
\hbox{\bf UCD-95-43} 
\hbox{\bf MADPH-95-918} 
%\hbox{\bf IUHET-299}
\hbox{November 1995}
\hbox{Revised: February 1999}
}$
}
%}

%
\medskip
\begin{center}
\bf
SEARCHING FOR INVISIBLE AND ALMOST INVISIBLE PARTICLES AT
$\bf\epem$ COLLIDERS 
\\
\rm
\vskip1pc
{\bf C.-H. Chen$^a$, M. Drees$^{a,b}$, and J.F. Gunion$^a$}\\
\medskip
\small\it
$^a$Davis Institute for High Energy Physics, 
University of California at Davis,\\
Davis, CA 95616, USA\\
$^b$Physics Department, University of Wisconsin, Madison, WI 53706, USA\\
\end{center}

\begin{abstract}
We explore the techniques, cross sections and expected signal
significance for detecting invisible and almost invisible
particles at LEP2 and the NLC by means of a hard photon tag. 
Examples from supersymmetry include
the lightest chargino and second lightest neutralino
when their masses are nearly the same as that of
the lightest neutralino (the LSP), and invisibly decaying sneutrinos. 
The importance of particular features of the detectors is discussed,
instrumentation for vetoing a fast $e^+$ or $e^-$ in the beam hole
being especially crucial.
\end{abstract}

\section{Introduction}

\indent\indent 
Models in which there are new particles that are either
themselves invisible, or that decay to invisible or nearly
invisible final states, abound in particle physics.
For example, one of the most popular and attractive models for physics
beyond the Standard Model (SM) is supersymmetry (SUSY).
The lightest supersymmetric particle (LSP, normally the lightest
neutralino, $\cnone$) of SUSY is invisible under the usual
assumption of $R$-parity conservation. Other (heavier) supersymmetric
partners of known particles (the sparticles) are usually
easily detectable at LEP2 and the next linear $\epem$ collider (NLC)
via events with energetic jets and/or leptons and missing energy,
when the fraction of energy carried
by the LSP and neutrinos in sparticle decays is not too large.
This is the case for the most popular model scenarios. However,
as outlined later, there exist SUSY models in which 
potentially visible jets/leptons from decays of
many and perhaps all the lower-mass sparticles
are either altogether absent or very to extremely soft. 
A heavy lepton doublet $(L^-,L^0)$ where the $L^0$ is stable in the
detector and $m_{L^-}-m_{L^0}$ is very small so that the $\ell^-$
from the $L^-\rta \ell^- \anti\nu L^0$ decay is very soft provides another
physics scenario which could be missed without employing a special
approach \cite{stokeretal}. Scenarios with extra scalar bosons have
been proposed in which some are invisible because of degeneracy \cite{GS}.
It is thus a matter of some urgency to explore
techniques, and to specify the required detector characteristics,
that will maximize our ability to detect invisible or `nearly invisible'
particles (denoted by IP and NIP, respectively)
{\it and to determine their mass}. If more than one IP and/or NIP
is present, the techniques we discuss could allow for
separation of the different signals and individual determination
of masses, depending upon statistics.

Even if a NIP decays to visible, but soft, particles, triggering on inclusive
NIP pair production is problematical. Typically, the mass difference
between the NIP and its invisible decay products must be
$\gsim 10\gev$ \cite{recentstudy} for inclusive 
pair production to be detectable in the presence of backgrounds.
For very small mass differences, the NIP would develop 
%Even if a NIP decays to potentially
%visible particles, and the mass difference involved is
%such that the decays yield visible tracks in the detector,
%triggering on inclusive pair production of NIP's
%by looking for events with soft secondary particles or jets
%and large missing energy is problematical. Typically, mass differences
%$\gsim 10\gev$ are required for the inclusive trigger to be
%effective in the presence of backgrounds leading to the same
%type of final state \cite{recentstudy}. And, 
%if the mass difference is $\lsim 0.5-1\gev$ the soft secondaries 
%are quite possibly invisible to the detector. As the mass difference
%decreases still further, the NIP would develop
a long enough lifetime that, if it is charged, 
a short track in the vertex detector
might be visible; the problem would be to trigger
on the event. Further, even if isolation of
signal events at the inclusive level is possible,
it could be very difficult to determine the mass
of the particles being produced via the usual spectrum endpoint
procedures. The possible solution to all these problems
is to require that a tagged photon be produced in association
with large missing energy from
the pair of IP's or NIP's. This is an idea that originated
for counting neutrinos \cite{GGR}, and has since been
employed for a number of entirely invisible SUSY sparticles, see for
example Refs.~\cite{KPWMGMSWMTATA,DDR}. In this paper we
emphasize photon tagging in the NIP context.

We focus particularly on some previously unexplored scenarios
in which the lightest chargino could be nearly invisible.
A common assumption regarding supersymmetry breaking
is that the gaugino masses are universal at the GUT scale, $\mgut$.
The renormalization group equations then
imply that the gaugino masses at scales below
a TeV are roughly related by $M_3\sim 3\times M_2$ and $M_2\sim 2\times M_1$,
where $M_3$ is the gluino mass and $M_2$ and $M_1$ are the $SU(2)$ and
$U(1)$ gaugino masses. Since the $\mu$ parameter
that also enters the chargino and neutralino mass matrices
is typically large, the above relations imply that the $\cnone$ 
LSP is mainly of the $U(1)$ bino variety with mass
of order $M_1$ and the lightest chargino, $\cpone$, is primarily
a wino with mass of order $M_2$.  This implies significant mass
splitting $\dmchi\equiv\mcpmone-\mcnone$, 
so that observation of $\cpone\cmone$
production in $\epem$ collisions is straightforward \cite{recentstudy}.
%ly accomplished
%by looking for $\ell\nu 2j+\etmiss$ and 
%$2\ell2\nu+\etmiss$ events
%arising from the $\cpmone\rta \cnone \ell\nu_{\ell},\cnone jj$  decays.
However, it is not impossible that the $\mgut$ boundary conditions 
are quite different and that $\dmchi$ could be small,
perhaps {\it very} small.
A small mass difference arises, in particular, in two cases:
\begin{description}
\item[(i)] High-$\mu$ scenario:
\footnote{In a separate paper \cite{modmodel} 
we shall explore in greater depth
a particular string model that leads to this scenario.}
if $M_2$ is substantially smaller than $M_1$, and $\mu\gg M_{1,2}$
then the $\cnone$ and $\cpone$ are both wino-like and closely
degenerate with $\mcnone\sim\mcpmone\sim M_2$. 
\item[(ii)] Low-$\mu$, large-$M_{1,2}$ scenario:
if $\mu\ll M_{1,2}$, then $\cnone$, $\cntwo$ and $\cpone$ are all
higgsino-like and nearly degenerate with masses $\sim\mu$.
\end{description}
Values of $\dmchi\lsim 10\gev$ (\ie\ in the problematical
region for normal inclusive detection) are not at all improbable
in either scenario, being almost automatic in scenario (i) and
requiring only $M_{1,2}\gsim 500 \gev$ in scenario (ii) (\ie\
well within the natural range for supersymmetry breaking).

The most challenging situation arises if 
$\dmchi$ lies below about a GeV,
since then the visibility of the $\cpmone$ comes into question.
In case (ii), $\dmchi\lsim 1\gev$
implies that $\mcntwo-\mcpmone$ and $\mcntwo-\mcnone$ will then
more or less automatically also be of order a GeV or less; in contrast,
in case (i) $\mcntwo$ is typically 
significantly larger than $\mcnone\sim\mcpmone$.
Extreme degeneracy ($\dmchi\lsim 1\gev$) 
between the lightest chargino and the LSP
can be achieved in case (i) for $M_2\gsim\mz/2$ if
$\mu\sim 1-2\tev$, whereas in case (ii) $M_{1,2}$ must be $\gsim 5\tev$
if $\mu \gsim \mz/2$.
%\footnote{We note that in neither case would this extreme degeneracy
%be removed by radiative corrections \cite{modmodel}.}
(Values of $\mcnone,\mcpone$ below $\mz/2$
are excluded by LEP1 limits on invisible/extra $Z$ 
decays \cite{modmodel}.)
Thus, only scenario (i) can remain technically natural for
$\dmchi\lsim 1\gev$, but from an experimental
perspective scenario (ii) also deserves exploration in this limit.

In this paper, we show that $\epem\rta \gam\cpone\cmone$
(and, in scenario (ii), $\epem\rta \gam\cnone\cntwo$) can
yield a viable signal at LEP2 and the NLC for the mass-degenerate
scenarios in question, depending upon
the value of the common mass (denoted $\mchitil$). 
If the soft secondary tracks
from $\cpmone$ decay are visible (with substantial
efficiency) then events containing a hard
photon and the visible $\cpmone$ remnants occur at a reasonable rate
so long as $\mchitil$ is not too close to the threshold allowed by
the required photon cut.
If the $\cpmone$ are effectively invisible, events of the type
$\epem\rta\gam+\etmiss$ will be adequately
enhanced (for appropriate cuts) over Standard Model (SM) expectations 
as to provide the required signal for $\mchitil$ up to somewhat
lower values. Whenever a signal is visible, at least
an approximate determination of $\mchitil$ will be possible.

%Before proceeding, we wish to note \cite{modmodel} that the $\cnone,\cpmone$
%[and $\cntwo$] cannot be lighter than $\lsim\mz/2$ in scenario (i) [(ii)].
%In scenario (i), $Z\rta\cpone\cmone$ decays would have been noticed
%either as an invisible width contribution or through an enhancement
%in the total $Z$ width.
%In scenario (ii), $\Gamma(Z\rta\cnone\cntwo)$
%alone is too large for consistency with experimental limits, with
%$\Gamma(Z\rta\cpone\cmone)$ providing additional inconsistency.
%These statements apply for masses up to within a fraction of a GeV of $\mz/2$.
%The $Z\rta \cnone\cnone,\cntwo\cntwo$ decays have small widths in these
%models, and do not provide useful limits.

\section{The Hard Photon Signals}

\indent\indent We will begin by focusing on the case in which
the light inos are effectively invisible so that the final
state is $\gam+\etmiss$. In practice, the only important signal processes 
are
\begin{enumerate}
\item $\epem\rta\gam\cpone\cmone$; 
\noindent and, in scenario (ii),
\item $\epem\rta\gam\cnone\cntwo$.

\noindent The only irreducible background is
\item $\epem\rta\gam\nu\overline\nu$.
\end{enumerate}
In both scenarios (i) and (ii) the $Z\cnone\cnone$
and $Z\cntwo\cntwo$ couplings are small, implying that $\gam\cnone\cnone$
and $\gam\cntwo\cntwo$ final states have negligible rate.
However, in scenario (ii) the $Z\cnone\cntwo$ coupling is maximal and reaction
(2) is important.
The background from $\epem\rta \gam\taup\taum$
in which {\it both} $\tau$'s decay to leptons or hadrons 
with small energy (say below a few GeV or so), 
or disappear down the beam pipe, is negligible by comparison
to reaction (3). A second, potentially very large, background is that from
$\epem\rta\epem\gamma$ events where neither the final $e^+$ 
nor $e^-$ is detected.
The techniques and experimental requirements for eliminating this
background are discussed below.
In our computations of the signal cross sections, we assume
that slepton and sneutrino exchange diagrams can be neglected.
In scenario (ii), this is automatically the case because the higgsino-like
$\cnone,\cpone,\cntwo$ have negligible coupling to $e\wtil e, e\snu$.
As described in Ref.~\cite{modmodel}, the specific string model
approach that leads to scenario (i) requires
a large $\mgut$-scale value of the soft supersymmetry breaking 
scalar mass parameter $m_0$,
implying that most sfermions (except possibly the lightest stop) 
have masses $\gsim 1\tev$, so that slepton
and sneutrino exchanges can again be neglected.  
Note that squarks,
sleptons and sneutrinos would then all be too heavy to be 
directly produced at the NLC,
let alone LEP2.  The only observable signals for SUSY at $\epem$
colliders would be those we now discuss.
               
We envision tagging the events using a photon that has substantial transverse
momentum.  For the study presented here, we require $p_T^\gam\geq \ptmin$,
with $\ptmin=10\gev$,
and $10^\circ\leq\thetagam\leq 170^\circ$, where $\thetagam$
is the angle of the photon with respect to the beam axis,
so as to guarantee that the photon enters a typical detector and
will have an accurately measured momentum. We define $\gam+\etmiss$
events by requiring that any other particle appearing in the $10^\circ$
to $170^\circ$ angular range must have energy less than $\emax$,
where $\emax$ is detector-dependent, but presumably
no larger than a few GeV. Kinematics can be used to show that
we can eliminate the $\epem\rta\epem\gam$ background
by vetoing events containing an $e^+$ or $e^-$ 
with $E>50\gev$ with angle $\thetamin\leq\thetae\leq 10^\circ$
with respect to either beam axis, or with $E>\emax$
and $10^\circ\leq\thetae\leq170^\circ$,
{\it provided} $\ptmin\gsim \sqrt s\sin\thetamin(1+\sin\thetamin)^{-1}$
(assuming $\emax$ is not larger than a few GeV). For $\ptmin=10\gev$,
this means that we must instrument the beam hole down to
$\thetamin=1.17^\circ$. In fact, instrumentation and vetoing
will be possible down to $\thetamin=0.72^\circ$ \cite{tbarklow},
implying that $\ptmin$ could be lowered to $\sim 6.2\gev$ without
contamination from the $\epem\rta\epem\gam$ background.
At LEP-192, beam hole coverage down to about $3.1^\circ$
is needed when using a $\ptmin=10\gev$ cut.

For ino masses above $\mz/2$, the key observation for reducing
the background reaction (3) and determining the ino mass
is to note that the missing mass
$\mzstar\equiv [(p^{e^+}+p^{e^-} - p^\gam)^2]^{1/2}$ 
can be very accurately reconstructed. For signals with
good overall statistical significance (in most cases $\nsd$, defined below,
$\gsim 5$ is adequate) one can plot events
as a function of $\mzstar$ and look for the threshold at $2\mchitil$
at which the spectrum starts to exceed the expectations from
$\gam\nu\anti\nu$.  We define an overall statistical significance 
$\nsd=S/\sqrt B$ for the signal by summing over all events with 
$\mzstar>2\mchitil$. Note, in particular, that
this cut eliminates the $Z$-pole contribution to the $\gam\nu\anti\nu$
background for $\mchitil>\mz/2$. In practice, one can often do better
(perhaps by $1\sigma$)
than this nominal $\nsd$ value by zeroing in on those $\mzstar$
bins with the largest deviations from $\gam\nu\anti\nu$ expectations.

\begin{figure}[htb]
\begin{center}
\centerline{\psfig{file=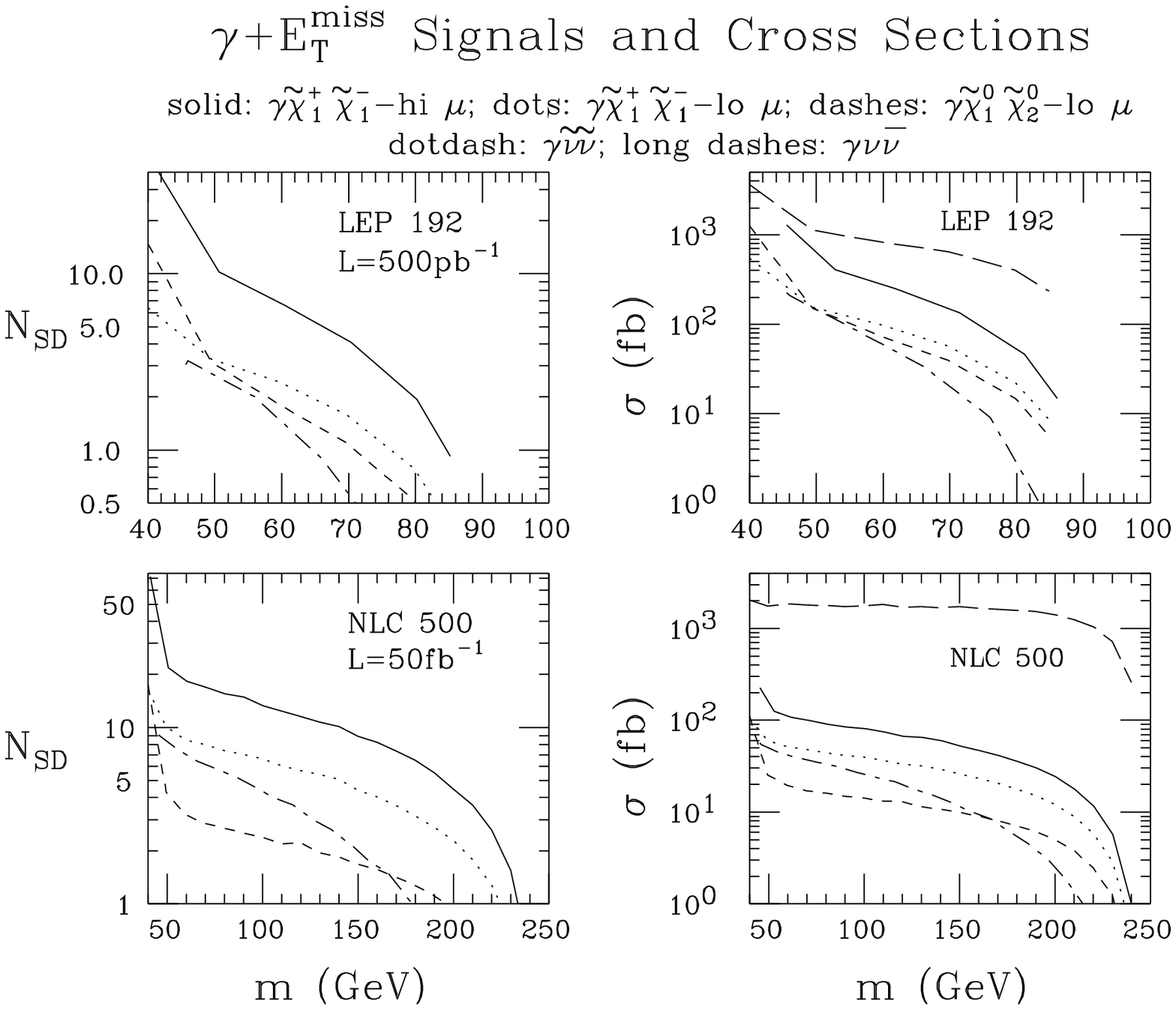,width=12cm}}
\bigskip
\begin{minipage}{11.5cm}       %%%% reduces width of caption to 10.5cm
\caption{We plot 
the statistical significance $\nsd=S/\protect\sqrt B$
in the $\gam+\etmiss$ channel 
as a function of NIP mass $m$ ($=\mchitil$ or $\msnu$).
%the $\epem\rta\gam\cpone\cmone$ signal in high-$\mu$ scenario (i);
%(dots)
%the $\epem\rta\gam\cpone\cmone$ signal in low-$\mu$ scenario (ii);
%(short dashes)
%the $\epem\rta\gam\cnone\cntwo$ signal in low-$\mu$ scenario (ii);
%and (dash dots)
%a signal from $\epem\rta \gam \snu\snubar$ in the limit of large
%slepton and ino masses. 
In all cases $B$ is computed from
$\epem\rta\gam\nu\anti\nu$ by integrating over $\mzstar\geq 2m$.
Results for LEP-192 (with $L=0.5\fbi$) and NLC-500 (with $L=50\fbi$)
are displayed. Also shown (right-hand panels) are the individual
cross sections for signals and background (long dashes). 
We employ the cuts: $p_T^\gam\geq 10\gev$;
$10^\circ\leq\thetagam\leq170^\circ$.
}
\label{eetogaminv}
\end{minipage}
\end{center}
\end{figure}

In Fig.~\ref{eetogaminv} we display our results.  For scenario (i),
the $\epem\rta\gam\cpone\cmone$ signal and $\epem\rta\gam\nu\anti\nu$
background cross sections are such as to yield (solid curves)
$N_{SD}=S/\sqrt B\geq 5$ for $\mchitil\lsim 65 \gev$ ($\lsim 200\gev$)
at LEP-192 (NLC-500) for total luminosities of $L=500\pbi$ ($50\fbi$),
respectively.  In contrast, 
for universal gaugino masses at $\mgut$ it is generally expected
that chargino pair production can be observed up to very nearly $\sqrt s/2$.
We note that the nominal $5\sigma$ signal observation requires
a small systematic uncertainty in our knowledge of the background,
given that $S/B \lsim 0.2$ ($\lsim 0.05$) at the $\mchitil$
value at LEP-192 (NLC-500) where $\nsd$ falls below 5.
%Systematic uncertainties in determining $p_T^\gamma$ in the electromagnetic
%calorimeter would aggravate the small $S/B$ problem.
Thus, it could be that the $\gam+\etmiss$ signal might not be viable
all the way out to the nominal $5\sigma$ mass value.

In scenario (ii), the $\cnone$, $\cpmone$ {\it and}
$\cntwo$ are all closely degenerate.
In this case, the magnitude of the $\gam+\etmiss$
signal depends upon whether or not we include $\epem\rta\gam\cnone\cntwo$
as a contribution.  We will present $\nsd$ values obtained 
for the $\gam\cpone\cmone$ and $\gam\cnone\cntwo$ channels
separately, keeping in mind that the mass degeneracy
means that in the $\gam+\etmiss$ channel they can be added together.
From the results presented in
Fig.~\ref{eetogaminv} we see that the $\gam+\etmiss$ signal
from the $\gam\cpone\cmone$ channel alone is much weaker in
this scenario where the $\cpmone$ are higgsino-like as
compared to the previous scenario where they are $SU(2)$-gaugino-like.
This is simply because the virtual $Z$-exchange contribution
is suppressed when the $\cpmone$ are higgsino-like (and
thus belong to a doublet vs. triplet SU(2) representation).
Without including the $\gam\cnone\cntwo$ channel LEP-192 (NLC-500)
can only achieve $\nsd\geq 5$ for $\mchitil\lsim 45\gev$ ($\lsim 140\gev$).
For the combined $\gam\cpone\cmone+\gam\cnone\cntwo$ channels
LEP-192 (NLC-500) can achieve $\nsd\geq 5$ for $\mchitil\lsim 55\gev$ 
($\lsim 170\gev$), better, but still not as large a reach
as for scenario (i).

By way of comparison, we also give results (dot-dashed curves)
for the $\gam+\etmiss$ signal deriving from $\epem\rta\gam\snu\snubar$ when
the $\snu$ decays invisibly to $\nu\cnone$ with 100\% branching
ratio. $BR(\snu\rta\nu\cnone)\sim 1$ is typical
of soft-SUSY-breaking models
having $\mhalf$ (the common gaugino mass) substantially
larger than $m_0$ (the common scalar mass) at $\mgut$; an example
is the very attractive Dilaton model with $m_0=\mhalf/\sqrt 3$ \cite{BGP}.
Detection of invisible sneutrinos in association with a photon tag was also
considered in Ref.~\cite{DDR}; there, several
other model contexts in which $BR(\snu\rta \nu\cnone)=1$ or is large 
are reviewed. Our procedures differ from those of Ref.~\cite{DDR}
in that we employ the $\mzstar\geq2\msnu$ cut 
to maximize the signal significance.  We conservatively
compute the signal in the approximation that charginos are heavy. 
We see from Fig.~\ref{eetogaminv} that a statistically
significant signal ($\nsd=5$) is not possible at LEP-192
for integrated luminosity of $L=500\pbi$, whereas $\msnu\lsim 100\gev$
could be probed at the NLC with $L=50\fbi$. 
(The signals found
in Ref.~\cite{DDR} at LEP-192 are also well below the $\nsd=5$ level.)
Finally, we note that in these models
the lighter $\stau$ eigenstate, $\stauone$, can be nearly
degenerate with the $\cnone$ (the crossover at $\mcnone=\mstauone$
often defines the upper limit on $\tanb$) in which case
$\gam\stauone\stauone$ production would provide the only viable
signature for the $\stauone$. 

For all the cases discussed above,
we have explored whether increasing the minimum $p_T^\gam$
required at a given $\mchitil$ or $\msnu$ would improve $\nsd$.
Even though $S/B$ can be improved at lower masses, the nominal $\nsd$ worsens
in all the cases examined. We also find that the distributions
of signal and background in $\thetagam$ are very similar 
(even in the $\gam\snu\snubar$ case)  so that additional 
$\thetagam$ cuts do not help.

The more limited range of viability for
the $\gam+\etmiss$ signals in scenario (ii) is a concern.
However, for $M_{1,2}$ values below a $\tev$ (but $\mu$ still
much smaller), the degeneracy among the
$\cnone,\cpmone,\cntwo$ will be only approximate and the leptons from
$\cpmone\rta\ell^{\pm}\nu\cnone$ or the photon from 
the one-loop decay $\cntwo\rta\gam\cnone$
(the decays $\cntwo\rta\ell^+\ell^-\cnone,q\anti q\cnone$ and
$\cntwo\rta\ell^{\mp}\nu\cpmone$ 
%followed by $\cpmone\rta\ell^{\pm}\nu\cnone$ 
usually having much smaller branching
ratio) would generally be visible. 

This leads us to consider the case in which
the $\cpmone$ and $\cntwo$ decay visibly, but the mass degeneracy is such that
the visible decay products are quite soft. 
Once again, the hard photon trigger will be crucial.
We have been unable to envision significant backgrounds to the
types of events in question (assuming the $\epem\rta\epem\gam$
events are vetoed, as described earlier).
In scenario (i) the signal events would contain a hard photon,
large missing energy,
and either two widely separated quasi-stable particle tracks, if the $\cpone$
has a long lifetime, or separated soft leptons, pions and/or jets.
In scenario (ii), the $\gam\cnone\cntwo$ events would have a single
extra soft isolated photon from the dominant 
$\cntwo\rta\gam\cnone$ decay. (Note that it 
would be possible to separate $\cpone\cmone$ from $\cnone\cntwo$ events.)
If the backgrounds to these two types of $\gam+\etmiss+\softvisible$
are truly negligible, then
it is the absolute rate obtained by combining all such events
that determines whether or not the events provide a viable signal. 

Let us first focus on the $\gam\cpone\cmone$ final state
in scenario (i).  The cross sections
of Fig.~\ref{eetogaminv} give event rates
that are sizeable for chargino masses substantially
above the values that can be probed by the $\gam+\etmiss$ signal.
%Crudely, the largest $\mchitil$ values
%that yield reasonable event rates are
%$\mchitil\sim 80$ at LEP-192 and $\mchitil\sim 240\gev$
%at NLC-500 for the $p_T^\gam\geq10\gev$ cut employed.
For our cuts,
we find about 25 (50) events at $\mchitil=80\gev$ ($240\gev$)
at LEP-192 with $L=500\pbi$ (NLC-500 with $L=50\fbi$).
With good efficiency either for detecting the $\cpmone$
as quasi-stable particle tracks in the vertex detector
or for detecting the $\cpmone$ decay products (\ie\ the soft pions, leptons
or jets), these event numbers may be adequate.
In scenario (ii), we find from Fig.~\ref{eetogaminv} [10,7] ([25,11])
events at $\mchitil=80\gev$ ($240\gev$)
for the $[\gam\cpone\cmone,\gam\cnone\cntwo]$
final states at LEP-192 (NLC-500).  These small numbers would appear
to be quite marginal; probably one would be able to extract
the signal obtained by combining the $\gam\cpone\cmone$
and $\gam\cnone\cntwo$ events only for masses $\lsim 75\gev$ ($\lsim 235\gev$)
at LEP-192 (NLC-500). A $5\sigma$ determination
of the two signal levels separately
would be possible only for a still more limited mass range.

\section{Lifetime and Branching Ratios for the $\bf \cmone$}

\begin{figure}[htb]
\begin{center}
\centerline{\psfig{file=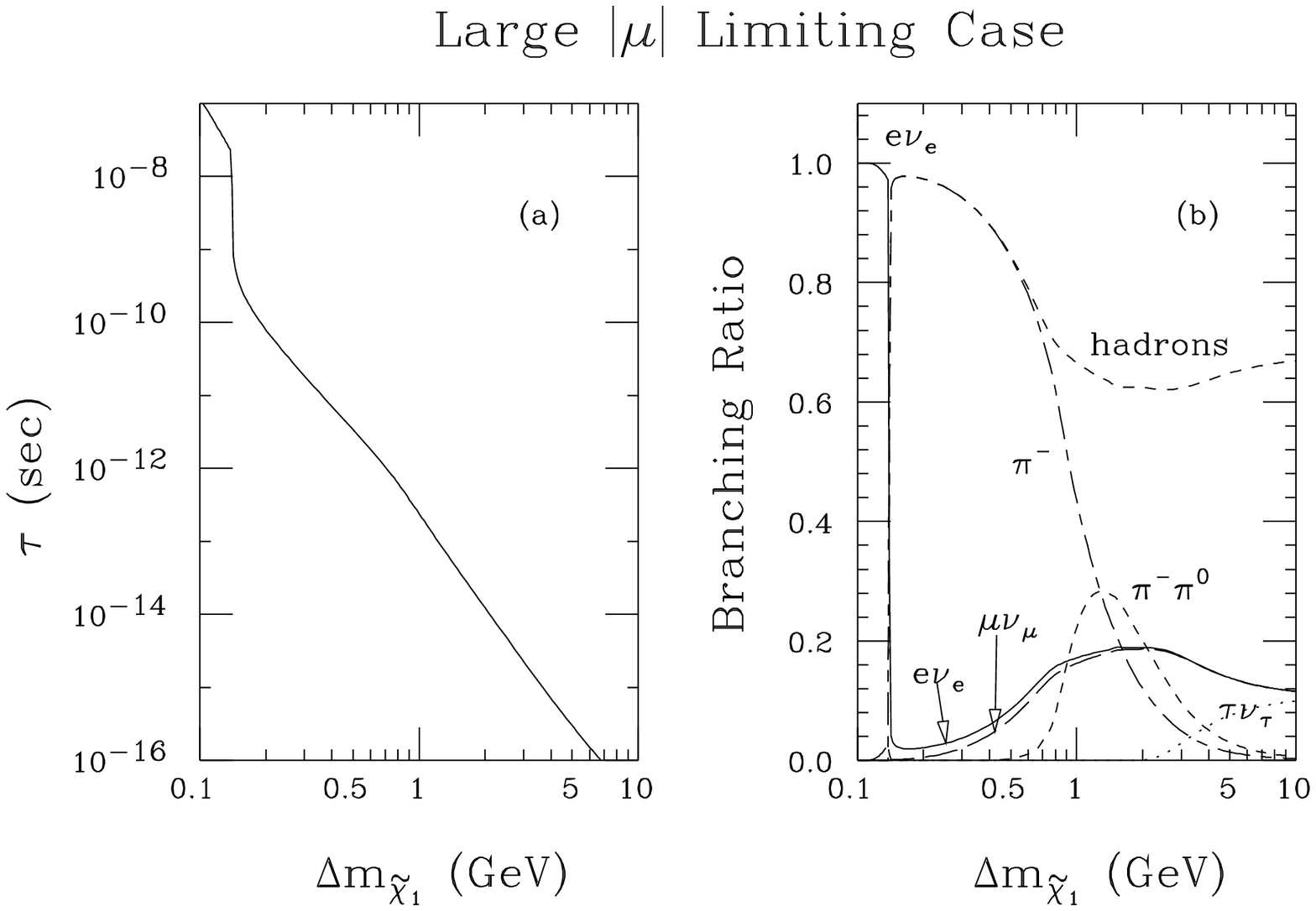,width=12cm}}
\bigskip
\begin{minipage}{11.5cm}       %%%% reduces width of caption to 10.5cm
\caption{We plot for scenario (i) 
the lifetime (a) and branching ratios (b) for the $\cmone$ as a function
of $\dmchi\equiv \mcpmone-\mcnone$. For $\dmchi< 1.5\gev$,
we explicitly compute and sum the $\pi^-\cnone$, $\pi^-\pi^0\cnone$,
$\pi^-\pi^0\pi^0\cnone$ and $\pi^-\pi^+\pi^-\cnone$ modes. These
merge into and are replaced by a computation of
the $q^\prime\overline q\cnone$ width for $\dmchi>1.5\gev$.
}
\label{lifebrs}
\end{minipage}
\end{center}
\end{figure}

In this section, we quantify the extent to which the lightest chargino
might or might not be visible. We have computed the branching ratios
and lifetimes for the $\cmone$. In Fig.~\ref{lifebrs}, we give explicit
results for the (more natural) large-$\mu$
scenario (i). For very small $\dmchi$,
$\cmone\rta\ell^-\nu_{\ell}\cnone$
($\ell=e,\mu$) is the only kinematically allowed decay mode.
%In case (ii), the corresponding $\mcntwo-\mcpone$ and $\mcntwo-\mcnone$
%mass differences will be such that $\cntwo\rta\ell^+\ell^-\cnone$,
%$\cntwo\rta\nu_{\ell}\overline\nu_{\ell}\cnone$ and 
%$\cntwo\rta\ell^{\mp}\nu_{\ell}\cpmone$, with subsequent $\cpmone$ decay,
%are the only allowed $\cntwo$ decays.
As the mass difference increases, $\cmone\rta\pi^-\cnone$ opens
up and remains dominant for $m_\pi<\dmchi\lsim 1\gev$.
Above that, the $\cnone\pi^-\pi^0$ and three-pion channels become
important. The sum of the one-, two- and three-pion channels merges
into $\cmone\rta q^\prime \overline q \cnone$ at $\dmchi\sim 1.5\gev$.  
For still larger
mass difference, $\cmone\rta\tau^-\nu_{\tau}\cnone$ 
becomes kinematically allowed. 
Details of the calculations will appear in
Ref.~\cite{modmodel}.  Here we simply note that the lifetime
and branching ratios are essentially independent of $\tanb$ and 
the sign of the $\mu$ parameter.  
%In Fig.~\ref{lifebrs}
%we have used a single curve for the
%$\pi^+\pi^0\cnone$ or $q^\prime\overline q\cnone$ branching ratio;
%the transition between the two different calculations of
%the hadronic mode decays is made at the matching point $\dmchi=2\gev$.
Finally, we note that scenario (ii) leads to very similar $\cmone$ branching
ratios, but, for a given $\dmchi$, about a 40\% longer lifetime.
And, we have already noted that in scenario (ii) the dominant $\cntwo$
decay is via one-loop graphs to $\gam\cnone$ for small mass differences.

Let us now consider implications for $\gam\cpone\cmone$ production.
Fig.~\ref{lifebrs}(a) shows that the produced $\cmone$
and $\cpone$ will travel distances of order
a meter or more (and thus appear as heavily-ionizing
tracks in the vertex detector and the main detector)
if $\Delta m_{\tilde{\chi}_1}<m_\pi$. 
For $m_\pi<\dmchi<1\gev$, $10~\mbox{cm}>c\tau> 100~\mu$m.
For $c\tau$ near 10 cm, the $\cpmone$ would
pass through enough layers of a typical vertex detector
that its heavily ionizing nature would be apparent.
For $c\tau$ in the smaller end of the above range, one would 
have to look for the single charged pion from the
dominant $\cpmone\to\pi^\pm\cnone$ mode. It emerges
at a finite distance of order $c\tau$ from the vertex
and would have momentum
$p_\pi\sim  \sqrt{\Delta m_{\tilde{\chi}_1}^2-m_\pi^2}$ in
the $\cpmone$ rest frame. The expected impact parameter resolution, $b_{\rm
res}$, of a typical vertex detector (we looked at the CDF Run II
vertex detector with the inner L00 layer in detail~\footnote{We thank
H. Frisch for providing details. The NLC vertex
detector can be built with similar characteristics (R. Van Kooten, private
communication). The innermost vertex detector at LEP is at $r=6.3$ cm,
implying less sensitivity there.}) as a function of momentum is such that
$c\tau/b_{\rm res}>3$ for $\dmchi<1\gev$, with quite large
values typical for $\dmchi<0.5\gev$. Such a high-$b$
pion in association with the $\gam$ trigger would constitute
a fairly distinctive signal. As discussed in the previous section, detection 
of any heavily-ionizing track and/or
{\it any} $\cpmone$ decay product would greatly enhance the significance
of the signal by removing the $\gam\nu\anti\nu$ background. 
For $\dmchi>1\gev$, the $\cpmone$ decay is prompt and one must
look for the soft leptons or hadrons emerging from the decay.
These might be difficult to detect if $\dmchi$ is not somewhat
larger. For instance, for $1<\dmchi<2\gev$, $\cpmone$ decays lead
to final states that are similar to those appearing in $\tau^\pm$ decays.
Without the hard photon plus $\etmiss$ tag, backgrounds to inclusive
$\epem\rta\cpone\cmone$ from $\gam\gam\rta \tau^+\tau^-$ might be
difficult to overcome, even if the chargino pair events can be triggered on.

\section{Final Remarks and Conclusions}

\indent\indent 
We have considered techniques for detecting particles
that decay invisibly
or nearly invisibly at $\epem$ colliders, focusing on the
implications for chargino detection
of scenarios in which $\dmchi\equiv\mcpmone-\mcnone$ is small,
including cases in which $\dmchi$
is neither small enough for the $\cpmone$
to produce a visible track in the detector nor large enough for the
leptons from $\cpmone\rta \ell\nu\cnone$ to have adequate momentum
to be visible. We have demonstrated that if the $\cpmone$
are effectively invisible, then $\epem\rta\gam+\etmiss$
events with $p_T^\gam\geq\ptmin=10\gev$ will be detectable
above the $\epem\rta\nu\anti\nu\gam$ background for a substantial
(but model-dependent) range of $\mcpmone$. 

In order that $\epem\rta\nu\anti\nu\gam$ be the primary background
in the $\gam+\etmiss$ channel, $\epem\rta\epem\gam$
events for which a fast final $e^+$ and $e^-$ are not seen must
be vetoed. At the NLC, for example, this implies that 
it is absolutely mandatory (and, apparently, straightforward)
for the detectors to have instrumentation in the 
$\thetamin$ to $10^\circ$ portion of the
beam hole, where $\thetamin\sim 1.17^\circ$ for $\ptmin=10\gev$.

We have delineated the lifetime and branching ratios of the $\cpmone$.
These can be used to determine the detector requirements
and machine environment that would alleviate the necessity for employing
the rather indirect $\gam+\etmiss$ signal for supersymmetry.
The hope is that one could observe
the $\gam\cpone\cmone$ events by tagging the photon, requiring
large $\etmiss$ and looking, in addition, for the
`quasi-stable' particle tracks and/or
the soft leptons or charged pions from the $\cpmone$ decays. 
We urge the detector
groups at LEP  and planning groups for the NLC to examine 
carefully the question of whether or not there is a band
in $\dmchi$ for which only the $\gam+\etmiss$ signature
(with the large $\gam\nu\anti\nu$ background) can be employed.
If tracks or remnants from the $\cpmone$ are visible with
good efficiency, we find that
the predicted rates for $\gam+\etmiss+\softvisible$ events
are such as to yield a viable $\gam\cpone\cmone$ signal
for $\mcpmone$ substantially nearer to the kinematic limit
implied by the photon trigger requirement than if
only the $\gam+\etmiss$ signature can be employed. 
In scenario (ii), similar statements
apply to $\epem\rta\gam\cnone\cntwo$ events,
where the soft leptons/pions are replaced by a single soft photon.

We stress that the $\gam+\etmiss$ and $\gam+\etmiss+\softvisible$
procedures are broadly applicable to isolating a signal for invisible
and nearly invisible particles.
The photon trigger also provides a general,
and, quite possibly, the only,  means for determining the
mass of any such particle.  
Mass determination is accomplished by employing
$\mzstar\equiv [(p^{e^+}+p^{e^-} - p^\gam)^2]^{1/2}$ to look
for the onset of signal events at $\mzstar$ equal to twice
the mass of the particle in question.  
%Although clearly easiest in the
%background-free $\gam+\etmiss+\softvisible$ mode, the 
%threshold in the $\mzstar$
%distribution relative to the $\gam\nu\anti\nu$ prediction is also visible
%for most cases with overall statistical significance of $5\sigma$ or better.
With good statistics, 
detection of several distinct $\mzstar$ thresholds can potentially
be used to separate signals appearing at different mass
scales due to different particles even when the associated events
are indistinguishable on the basis of event characteristics.

In Ref.~\cite{modmodel}, we explore Tevatron and LHC
detection of gluinos when $\mgl$ is near $\mcpmone\sim\mcnone$.
Despite the softness of the jets in $\gl\rta q^\prime\anti q\cpmone,q\anti
q\cnone$ decays and the invisibility of the $\cpmone$ decay products, 
we find that detection of gluino pair events in $jets+\etmiss$
final states will still be possible for much the same mass ranges as before.
This is because gluino pairs have a high probability
of being made in association with one or more energetic jets. 
Thus, both lepton and hadron machine data 
will allow us to probe supersymmetry even when mass splittings
among the light supersymmetric particles are small.

\section{Acknowledgements}

This work was supported in part by U.S. Department of Energy grants
DE-FG03-91ER40674 (JFG, CHC) and DE-FG02-95ER40896 (MD), 
the Davis Institute for High Energy Physics, the Wisconsin Research Committee
using funds granted by the Wisconsin Alumni Research Foundation (MD),
and by a grant from the Deutsche Forschungsgemeinschaft under
the Hiesenberg program (MD).
We thank T. Barklow and U. Nauenberg for helpful discussions and information.

\clearpage
 
\end{document}